\documentclass[a4paper,11pt]{article}
\pdfoutput=1 

\usepackage{jheppub} 

\usepackage[T1]{fontenc} 
\usepackage{amsmath}
\usepackage{amsfonts}
\usepackage{amssymb}
\usepackage{graphicx}
\usepackage{hyperref}
\usepackage{color}
\usepackage{caption}
\usepackage{verbatim}
\usepackage[normalem]{ulem}
\usepackage{subfigure}
\usepackage{slashed}
\usepackage{cancel}
\usepackage{braket}
\usepackage{multicol}
\usepackage{tcolorbox}
\tcbuselibrary{theorems}
\usepackage{bm}
\usepackage{breqn}

\usepackage{physics}
\usepackage{tikz-feynman}
\usepackage{xfrac}

\def\micro{{\tt micrOMEGAs}}

\def\darkON{{\tt darkOmegaN}}
\def\darkO{{\tt darkOmega}}
\def\exclu{{\tt Excluding2010}}

\title{\boldmath Coscattering in the Extended Singlet-Scalar Higgs Portal}

\author[a,b]{Bastián Díaz Sáez,}
\author[c]{Jayita Lahiri,}
\author[d]{Kilian Möhling}
\affiliation[a]{Instituto de Física, Pontificia Universidad Católica de Chile,
Avenida Vicuña Mackenna 4860, Santiago, Chile}
\affiliation[b]{Millennium Institute for Subatomic Physics at the High-Energy Frontier (SAPHIR),
Fernández Concha 700, Santiago, Chile}
\affiliation[c]{II. Institut für Theoretische Physik, Universität Hamburg, Luruper Chaussee 149, 22761 Hamburg,
Germany}
\affiliation[d]{Institut für Kern- und Teilchenphysik, TU Dresden, Zellescher Weg 19,
01069 Dresden, Germany}
\emailAdd{bastian.diaz@uc.cl}
\emailAdd{jlahiri29@gmail.com}
\emailAdd{kilian.moehling@tu-dresden.de}

\abstract{We study the coscattering mechanism in a simple Higgs portal which add two real singlet scalars to the Standard Model.
In this scenario, the lighter scalar is stabilized by a single $\mathcal{Z}_2$ symmetry and acts as 
the dark matter relic, whose freeze-out is driven by conversion processes. The heavier scalar becomes an unstable state which participate actively in the coscattering. We find viable parameter regions fulfilling the measured relic abundance, while evading direct detection and big-bang nucleosynthesis bounds. In addition, we discuss collider prospects for the heavier scalar as a long-lived particle at present and future detectors.}

\begin{document} 
\maketitle
\flushbottom

\section{Introduction}
Coscattering \cite{DAgnolo:2017dbv} or Conversion-driven freeze-out \cite{ Garny:2017rxs} is a thermal dark matter (DM) framework in which the dark matter relic abundance is determined by the freeze-out of inelastic conversions in the dark sector.
In the typical coannihilation regime such processes are assumed to be rapid enough to keep the dark sector in chemical equilibrium (CE) even long after the freeze-out of the DM from the thermal plasma. In contrast, in the coscattering scenario the dark sector falls out of CE roughly once the conversion rates drop below the Hubble expansion rate. For previous studies of this mechanism in several different models see Refs. \cite{Cheng:2018vaj, Garny:2018icg, DAgnolo:2018wcn, Brummer:2019inq, Junius:2019dci, DAgnolo:2019zkf, Garny:2021qsr, Herms:2021fql, Heeck:2022rep, Filimonova:2022pkj, Heisig:2024xbh}.  

A typical feature of the coscattering regime is the presence of long-lived particles (LLP), because the 
small coupling strength between the relic and unstable dark partner required for a fast freeze-out of the conversions in turn implies a narrow decay with of the dark partner. 
Furthermore, the masses of the DM species must be highly degenerate, $\Delta m \ll m_{DM}$, as otherwise the Boltzmann suppression of the conversion rate leaves the coscattering mechanism inactive.
Since the LLPs can couple much more strongly to the SM they are excellent candidates for direct detection of DM at present or future colliders \cite{Curtin:2018mvb, Cottin:2024dlo, Heisig:2024xbh}.

In this paper, we study the coscattering mechanism in one of its perhaps simplest possible realizations,
a two real singlet-scalar model coupling to the SM through the Higgs portal \cite{Ghorbani:2014gka,Casas:2017jjg}. Here, the lighter scalar is stabilized by a discrete $\mathcal{Z}_2$ symmetry, while the second scalar 
acts as the unstable dark partner. We find that the coscattering regime allows for DM masses at the
EW scale, while the dark partner constitutes a LLP with $c\tau \lesssim 10^5$ km.\newline

The paper is structured in the following way. In Sec. \ref{sec1} we present the model. In Sec. \ref{sec2}
we discuss the calculation of the relic abundance in its different regimes, paying special attention to the coscattering regime.
In Sec. \ref{sec3} we present the relevant experimental constraints and obtain results for the expected
lifetimes of the LLP. Finally, we give some concluding remarks in Sec. \ref{sec4}.

\section{Model}\label{sec1}
 We consider the SM extended by two real singlet-scalars $S_1$ and $S_2$.
 $S_1$ is taken to be the lighter scalar, which is stabilized by a $Z_2$ symmetry under which $S_i \rightarrow -S_i$, while the SM fields transform trivially \cite{Casas:2017jjg, Ghorbani:2014gka}. 
 As a result, the scalars couple to the SM only via the Higgs field. 
 The corresponding Lagrangian in the scalar mass basis (for more details see App. \ref{app1})
 is given by
 \begin{align}\label{pot}
     \begin{split}
        \mathcal{L} = \mathcal{L}_{SM} &+ \sum_{i=1,2}\left(\frac{1}{2}(\partial_\mu S_i)^2 - \frac{m_i^2}{2} S_i^2 - \lambda_{i4}S_i^4\right) - \lambda_{22}S_1^2S_2^2 - \lambda_{13}S_1S_2^3 - \lambda_{31}S_1^3S_2\\
        & - \left(\lambda_{H1}S_1^2 + \lambda_{H2}S_2^2 + \lambda_{12}S_1S_2\right)\left(|H|^2 - \frac{v_h^2}{2}\right),
     \end{split}
 \end{align}
where $H$ denotes the SM Higgs doublet and $v_h\approx 246$ GeV the Higgs vacuum expectation value (vev).
None of the new scalars acquire a vacuum expectation value. 

In the following, we consider $(m_1,m_2,\lambda_{H1}, \lambda_{12}, \lambda_{H2}, \lambda_{22})$ the set of independent model parameters and denote the mass difference between the scalars by $\Delta m \equiv m_2 - m_1 > 0$. In the coscattering regime, the couplings $\lambda_{13}$ and $\lambda_{31}$ play a similar
role to $\lambda_{22}$ and are omitted for simplicity.

\section{Coscattering or Conversion-driven freeze-out}\label{sec2}
In the coscattering regime we explicitly do not assume CE within the dark sector during the evolution of the DM number densities $n_i$ up to the point of freeze-out. 
As a result, the full coupled Boltzmann equations (cBE),
assuming all possible interaction terms, have to be solved in order to obtain the correct DM relic abundance. In the following we introduce $x = m_1/T$ together with the typical definition of the
DM yield $Y_i := n_i/s$, where $s$ denotes the entropy density. 

\subsection{Boltzmann equations}
The cBE for $Y_1$ and $Y_2$ reads
\begin{subequations}\label{beq}
    \begin{align}
        \begin{split}
         \frac{dY_1}{dx} &= \frac{1}{3H}\frac{ds}{dx}\bigg[\ev{\sigma_{1100} v}\left(Y_1^2 - Y_{1e}^2\right) + \ev{\sigma_{1200} v}\left(Y_1Y_2 - Y_{1e}Y_{2e}\right) \\ &+ \ev{\sigma_{1122} v}\left(Y_1^2 - Y_{2}^2\frac{Y_{2e}^2}{Y_{1e}^2}\right)  + \frac{\Gamma_{1\rightarrow 2}}{s}\left(Y_1 - Y_2\frac{Y_{1e}}{Y_{2e}}\right) + \frac{\Gamma_2}{s}\left(Y_2 - Y_1\frac{Y_{2e}}{Y_{1e}}\right)\bigg],
        \end{split} \\
        \begin{split}
          \frac{dY_2}{dx} &= \frac{1}{3H}\frac{ds}{dx}\bigg[\ev{\sigma_{2200} v}\left(Y_2^2 - Y_{2e}^2\right) + \ev{\sigma_{1200} v}\left(Y_1Y_2 - Y_{1e}Y_{2e}\right) \\ &- \ev{\sigma_{1122} v}\left(Y_1^2 - Y_{2}^2\frac{Y_{2e}^2}{Y_{1e}^2}\right)  - \frac{\Gamma_{1\rightarrow 2}}{s}\left(Y_1 - Y_2\frac{Y_{1e}}{Y_{2e}}\right)-\frac{\Gamma_2}{s}\left(Y_2 - Y_1\frac{Y_{2e}}{Y_{1e}}\right)\bigg],
        \end{split}
    \end{align}    
\end{subequations}
where $H$ denotes the Hubble rate, $0$ stand for any SM particles, and $1,2$ for $S_1$ and $S_2$ respectively. The equilibrium yields are given by
\begin{subequations}
    \begin{align}
        Y_{1e}(x) &= \frac{45}{4\pi^4}\frac{x^2}{g_{*S}(x)}K_2(x), \\
        Y_{2e}(x) &= \frac{45}{4\pi^4}\frac{x^2}{g_{*S}(x)}\frac{m^2_2}{m^2_1}
        K_2\left({\textstyle \frac{m_2}{m_1}} x\right),
    \end{align}
\end{subequations}
where $K_2(x)$ is the modified Bessel function of the second kind, $g_{*S}(x)$ the number of effective degrees of freedom associated to the entropy density $s = \frac{2\pi^2}{45} g_{*S}(T) T^3$.
In contrast to the cBE for coannihilation, eqs.~\eqref{beq} explicitly contains the DM conversion rate
\begin{eqnarray}
 \Gamma_{1\rightarrow 2} = \sum_{k,l}\ev{\sigma_{1k\rightarrow 2l} v} n_{k,e},
\end{eqnarray}
where $k$ and $l$ denote light SM states. The calculation of the relevant conversion cross sections together with their thermal average is presented in App. \ref{App:conversion-rate}. 
The second important conversion process is given
by decays of the unstable partner $S_2$ with the thermally averaged decay rate \cite{Garny:2017rxs}
\begin{eqnarray}\label{decayeq}
 \Gamma_{2} \equiv \frac{K_1(m_2/T)}{K_2(m_2/T)} \sum_X \Gamma(2\to X).
\end{eqnarray}
We solve the above cBE using \micro \ 5.3.41 \cite{Belanger:2001fz, Alguero:2022inz}, considering three separate sectors: i) the SM, ii) the DM candidate $S_1$, and iii) a dark sector for $S_2$. \micro \ solves all the relevant average cross sections, including the two and three-body decay widths of $S_2$ considering Lorentz time effects. To quantify the impact of coscattering and compare the results obtained from the full cBE to
the results assuming CE we use \cite{Alguero:2022inz}
\begin{eqnarray}\label{delta1s}
 \Delta_{1s}^\Omega \equiv 1 - \frac{\Omega h^2 (\text{1 sector})}{\Omega h^2 (\text{2 sectors})}, 
\end{eqnarray}
where $\Omega h^2$(1 sector) is obtained using the \darkO \ function of \micro \ and $\Omega h^2$(2 sectors) is obtained from \darkON \footnote{In the present paper we did not make explicit use of the function $\Delta_{2s}^\Omega$ defined in \cite{Alguero:2022inz}, although part of the analysis in this section contemplates the information that could be obtained with that function.}. The scaling of each process with the model parameters are listed in Table \ref{tab1}.

\begin{table}
\begin{center}
\begin{tabular}{||c c c||} 
 \hline
 Initial & Final & Scaling  \\ [0.5ex] 
 \hline\hline
 1 1 & 0 0 & $\lambda_{H1}^2, \lambda_{12}^2$  \\ 
 \hline
 2 2 & 0 0 & $\lambda_{H2}^2, \lambda_{12}^2$  \\
   \hline
  \hline
 1 1 & 2 2 & $\lambda_{H1}^2, \lambda_{12}^2, \lambda_{H2}^2, \lambda_{22}^2$   \\
 \hline
 \hline
  1 2 & 0 0 & $\lambda_{H1}^2,\lambda_{12}^2, \lambda_{H2}^2$  \\
 \hline
 1 0 & 2 0 & $\lambda_{H1}^2,\lambda_{12}^2, \lambda_{H2}^2$   \\
 \hline
 2 & 1 0 & $\lambda_{12}^2$   \\  
 \hline
\end{tabular}
\caption{Scattering and decay processes with their corresponding scaling, ignoring the quartic couplings $\lambda_{13}$ and $\lambda_{31}$.}
\label{tab1}
\end{center}
\end{table}

\subsection{Relic abundance}\label{relicsub}

The basic characteristic of the conversion-driven freeze-out in the two scalar Higgs portal are:
\begin{enumerate}
 \item $S_1$ remains in CE with $S_2$ only through either (inverse) decays or coscattering processes $10\leftrightarrow 20$. 
 \item Annihilation processes $S_1 S_i \rightarrow XX$ involving $S_1$ can be neglected. 
\end{enumerate}

The first condition requires that $\lambda_{12}$ is non-vanishing but small enough for the conversion processes not to surpass the Hubble expansion rate at $T \lesssim m_1$. On the other hand, to prevent an early freeze-out and overabundance of DM it is required that $S_2$ couples strongly with the Higgs $\lambda_{H2} \sim 1$.
The second condition is fulfilled only when in addition to $\lambda_{12}$, also $\lambda_{H1}\ll 1$.
In case of on-shell (inverse) decays of $S_2$, the dark sector can stay in CE for much smaller couplings compared to the case of off-shell decays, however, we have checked that 
in both cases conversion-driven freeze-out is possible (in contrast to \cite{ DAgnolo:2017dbv} who assumed that 2-body decays are forbidden).
In the last part of this section we analyse this point in more detail. Lastly, we note that the contact interaction terms in eq. \eqref{pot} can not be arbitrarily large, as otherwise they will recover CE between $S_1$ and $S_2$. 
The impact of $\lambda_{22},\lambda_{13}$ and $\lambda_{31}$ is discussed in more detail at the end of this section. In Table~\ref{tab1} we show the parameter dependence for each process that enters in eqs.~\ref{beq}. \newline 

\begin{figure}[t!]
\centering
\includegraphics[width=0.95\textwidth]{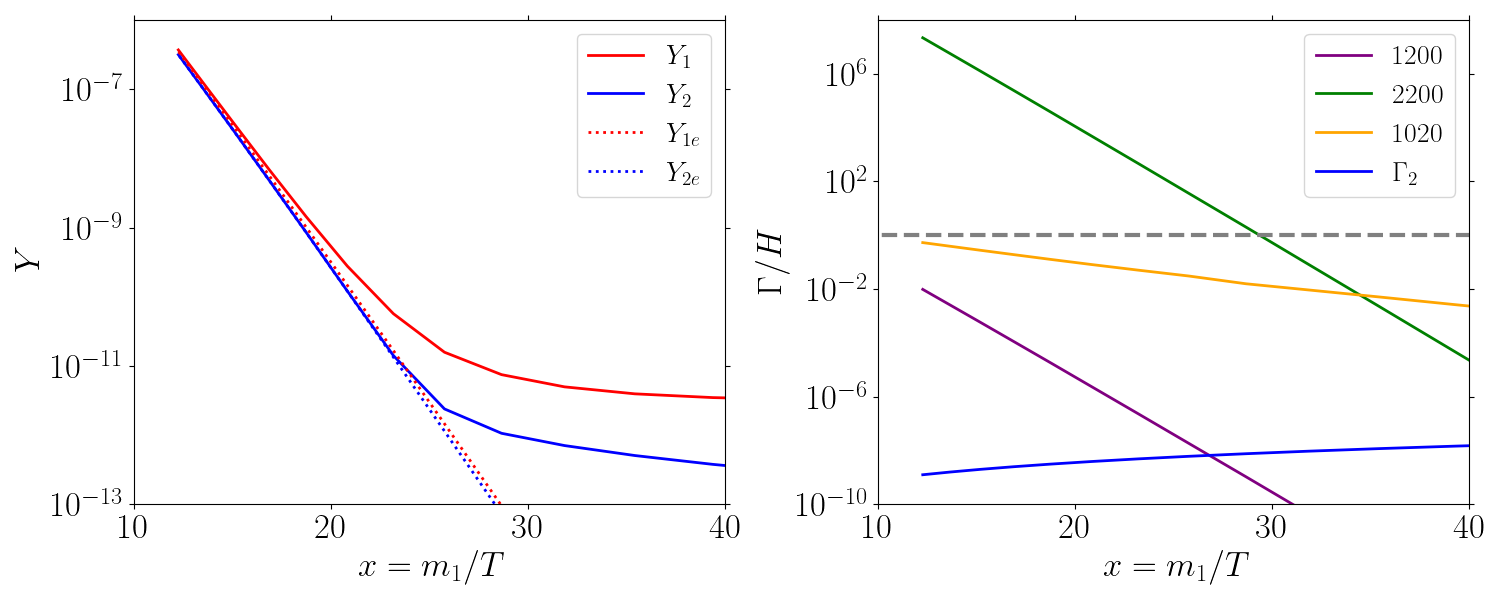}
\caption{(left) Relic abundance in the coscattering regime for the benchmark point $(m_1, m_2) = (500, 505)$ GeV, $(\lambda_{12}, \lambda_{H2}) = (2.6\times 10^{-5}, 1)$. (right) Scattering and decay rates compared to the Hubble rate as a function of the inverse temperature at the benchmark point. The dashed horizontal line represents when the rate interactions equal the Hubble rate.} 
\label{plot1}
\end{figure}

In order to simplify the discussion and exploration of the parameter space of the model in the coscattering framework, we define the \textit{simplest benchmark scenario} (SBS) considering $\lambda_{H1} = \lambda_{22} = 0$, and the relevant parameters as
\begin{eqnarray} (m_1, m_2,\lambda_{H2},\lambda_{12}).
\end{eqnarray} 
Deviations from the SBS will be explicitly shown in some parts of the paper. As a warm up example of the features of coscattering in the SBS, in Fig. \ref{plot1} we show a typical evolution of the DM yield in the coscattering regime fulfilling the correct relic abundance $\Omega h^2 = 0.12$, for $(m_1,m_2) = (500,505)$ GeV, and $(\lambda_{12}, \lambda_{H2}) = (2.6\times 10^{-5}, 1)$. Notice that $Y_1$ deviates from its equilibrium already near $x \approx 12$, whereas $Y_2$ stays in equilibrium for longer. This behavior is characteristic of the 
coscattering regime.
In the right plot, we compare the reaction rates with the Hubble expansion, where 
$\Gamma_{ij} \equiv \frac{\gamma_{ij\rightarrow kl}}{n_{1e}}$ and $\gamma_{ij\rightarrow kl}$ 
denotes the reaction density. In particular it can be seen that the DM conversion rate $1020$ (yellow line) drops below
the Hubble expansion at the same time as $S_1$ starts to freeze out from the thermal bath.
Note that in the SBS scenario, decays and coannihilations are well below the Hubble rate and are completely negligible during the freeze-out process.

\begin{figure}[t!]
\centering
\includegraphics[width=0.7\textwidth]{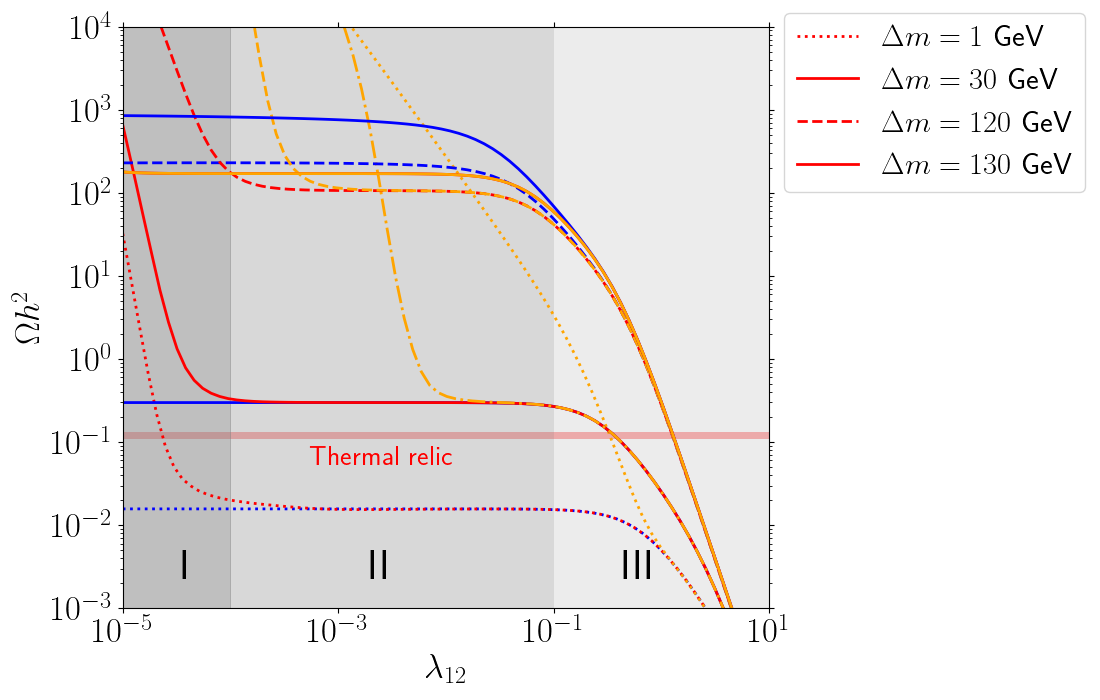}
\caption{Relic abundance obtained in the SBS considering $m_1 = 500$ GeV. The red curves are obtained with \darkON , the blue ones with \darkO, and the orange ones without considering the process 1020 in eqs.~\ref{beq} (in \micro \ this quantity can be obtained using the command "\exclu"). Note that the solid 
red curve is covered by the solid orange curve. The regions shown here as I, II and III correspond to the case of $\Delta = 30$ GeV.}
\label{plot2}
\end{figure}

With this simple picture in mind, we now vary $m_2$ and $\lambda_{12}$ and study their impact on the
relic abundance. We have performed a grid scan over $\lambda_{12}\in [10^{-5},10]$ and 
$m_2\in [500, 630]$ GeV, keeping $m_1 = 500$ GeV and $ \lambda_{H2} = 1$ fixed.
The results are shown in Fig.~\ref{plot2}, where the red curves correspond to the solutions of the
full cBE obtained with \darkON, the blue curves where obtained using \darkO\ and
the orange curves where obtained ignoring the conversion processes $1020$.
While the relic abundance shows a similar behavior when varying $\lambda_{12}$ for different values 
of $\Delta m$, the predicted relic abundance differs very strongly. 
This is due to the fact that the effective annihilation rate determining the point of freeze-out
$e^{-2x\Delta m/m_1} \ev{\sigma_{2200} v}$ is exponentially suppressed for large $\Delta m$. This suppression leads to a smaller effective cross section which implies a
faster freeze-out and larger relic abundance, as can be seen in Fig. \ref{plot2}.

As an example to better understand the dependence of $\Omega h$ on $\lambda_{12}$, we consider
the case $\Delta m=30$ GeV (dot-dashed line). In Fig. \ref{plot2} we have highlighted three 
distinct regions for the behaviour of the relic abundance.
The coscattering mechanism is only active in region I where $\lambda_{12}$ is small enough so that the $1020$ conversion processes freeze-out quickly. 
As the coupling increases, CE is recovered and the relic abundance becomes insensitive to $\lambda_{12}$
in region II. In this case the relic abundance is mainly determined by $S_2$ annihilation, which is also called \textit{mediator freeze-out regime} \cite{Junius:2019dci}. Finally, in region III for $\lambda_{12} \gtrsim 0.1$ coannihilations between $S_1$ and $S_2$ become relevant and the relic abundance again depends on $\lambda_{12}$. 

For each $\Delta m$ in Fig.~\ref{plot2} we have also included the corresponding relic abundance 
obtained from \darkON \ when neglecting the processes 1020, but keeping decays\footnote{In \micro\ this
is achieved using the option \exclu.}. The resulting abundances are plotted as the orange lines, highlighting the fact that for small values of $\lambda_{12}$ decays are not able to support CE in the absence of processes of the type 1020. 
In the case of on-shell decays (solid orange), where the decay rates are much larger, CE is maintained also at small $\lambda_{12}$. 
In this case the orange and red lines overlap in the whole range of small couplings.
We have also included the results for the relic abundance calculated using the function \darkO \ of \micro \ (blue lines). This function assumes that that CE between $S_1$ and $S_2$ is maintained during the entire evolution of the DM yield. In case of $\Delta m = 1$ and $30$ GeV, the relic abundance obtained with the functions \darkO \ and \darkON \ agree very well in regions II and III, indicating that CE is present. In case of $\Delta m = 120$ GeV, the results assuming
CE are larger by roughly a factor of two, which further increases for larger mass differences.
We have checked that in these cases the rate of (inverse) decays remains above the Hubble expansion, 
ensuring CE. The correct relic abundance is therefore obtained from \darkO,
while \darkON \ assumes separate CE of the different sectors, which is unrealistic, particularly when on-shell decays are present.

\begin{figure}[t!]
\centering
\includegraphics[width=0.45\textwidth]{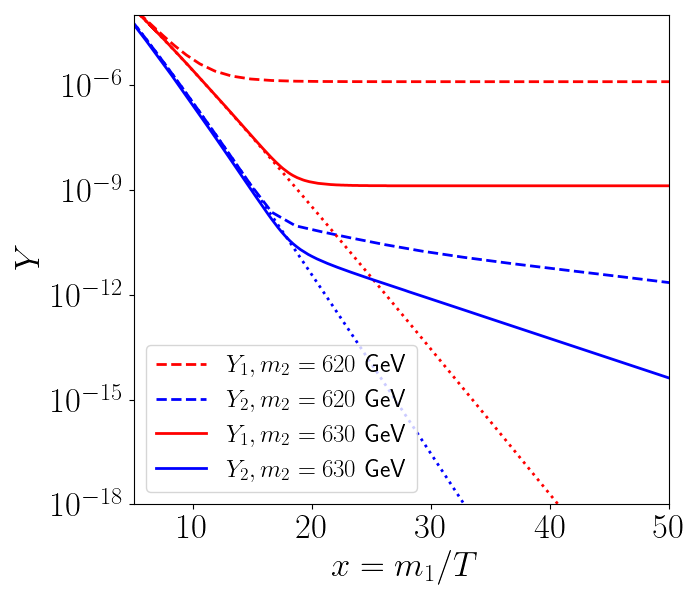}\quad
\includegraphics[width=0.45\textwidth]{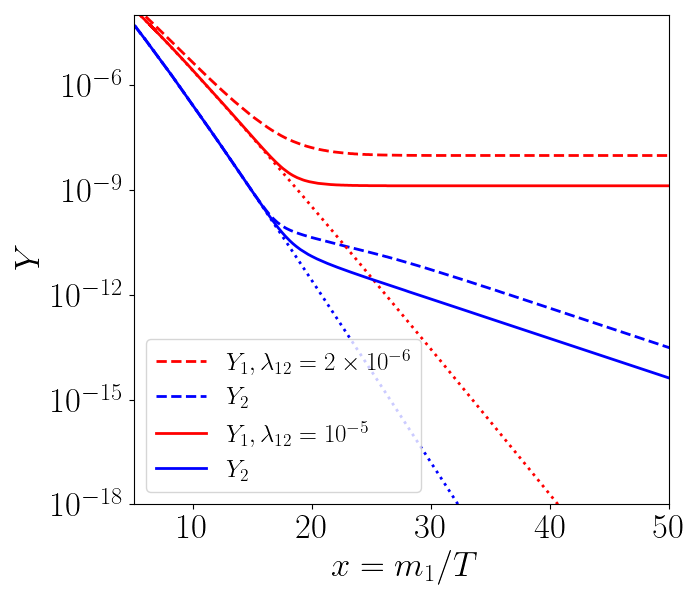}
\caption{(left) DM yield evolution for the case of off-shell (dashed) and on-shell (solid) decays with
$m_1 = 500$ GeV, $\lambda_{12} = 10^{-5}$, $\lambda_{H2} = 1$, and $\lambda_{H1} = \lambda_{22} = 0$. (right) DM yield evolution for on-shell decays for $(m_1,m_2) = (500, 630)$ GeV, with $\lambda_{12}=2\times 10^{-6}$ (dashed)
and $\lambda_{12}=10^{-5}$ (solid).} 
\label{plot3}
\end{figure}

From the case with $m_2 = 630$ GeV shown in Fig.~\ref{plot2}, we have seen that on-shell decays maintain CE for much smaller $\lambda_{12}$ values. To further illustrate this fact, in Fig.~\ref{plot3}~(left) we show the early departure from CE of the yield $Y_1$ in the off-shell case, $m_2 = 620$ GeV, compared to the on-shell case, $m_2 = 630$ GeV, for a fixed value of $\lambda_{12}$. In Fig.~\ref{plot3}~(right) we show that, once on-shell decays are present, small enough values of $\lambda_{12}$ also break the CE\footnote{We have checked this in the case of on-shell decays, coscattering appears in the ballpark of $\mathcal{O}(\lambda_{12}) \propto 10^{-6}$, but as a strong overabundance is obtained in this parameter space region, we do not focus on this case in this paper.}. In other words, as $\Delta m$ increases, the relevance of decays in maintaining CE becomes stronger, requiring smaller $\lambda_{12}$ for coscattering. 

\begin{figure}[t!]
\centering
\includegraphics[width=0.6\textwidth]{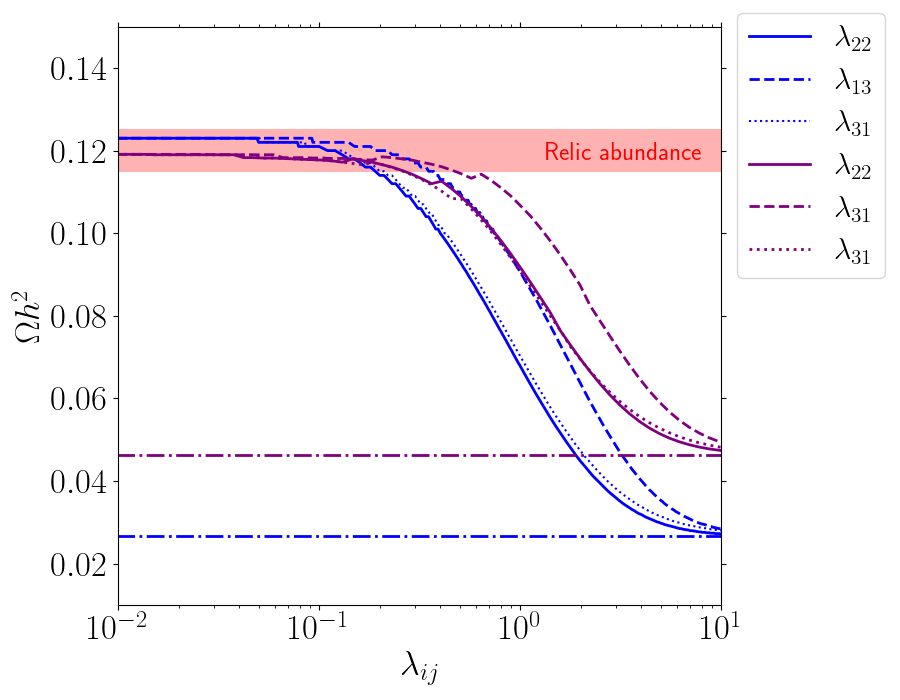}\quad
\caption{Relic abundance behavior as a function of the contact term couplings $\lambda_{22}, \lambda_{13}$ and $\lambda_{31}$. The solid, dashed and dotted lines are obtained with \darkON\ , whereas the dashed-dot lines are obtained with \darkO. The blue lines correspond to $(m_1, m_2) = (500, 505)$ GeV and $(\lambda_{H1}, \lambda_{12}, \lambda_{H2}) = (0, 2.6\times 10^{-5}, 1)$, whereas the purple lines correspond to $(m_1, m_2) = (500, 510)$ GeV and $(\lambda_{H1}, \lambda_{12}, \lambda_{H2}) = (0, 3.2\times 10^{-5}, 1)$. In each case, $\lambda_{12}$ was fixed to obtain the correct relic abundance (red band) with \darkON\ for vanishing contact couplings.} 
\label{plot4}
\end{figure}

On the other hand, the contact terms proportional to the couplings $\lambda_{22}, \lambda_{13}$ and $\lambda_{31}$ can have a strong 
impact on the relic density. In particular, when they take sizable values, i.e. either $\lambda_{22}, \lambda_{13}$ or $\lambda_{31} \gtrsim 0.1$, they 
bring $S_1$ and $S_2$ back into CE. To quantify this, in Fig.~\ref{plot4}~(left) we show the effect of each separate coupling on the relic abundance for two set of masses. In each case, the value of $\lambda_{12}$ was fixed in order to obtain the correct relic abundance by \darkON \ in the limit of vanishing contact terms. As the contact couplings get sizable values, they start to affect the relic abundance calculation with \darkON\, as they tend to establish partial CE between $S_1$ and $S_2$. Once the contact couplings are big enough, the CE is established, such that the calculation using \darkO\ and \darkON\ agree with each other\footnote{It is interesting to remark that for the case in which $\lambda_{22}$ takes sizable values, and $\lambda_{12}$ remains sufficiently small to not maintain CE between $S1$ and $S_2$, one recover the yield dynamic of two stable DM, known as assisted freeze-out \cite{DiazSaez:2021pfw} (also see \cite{Belanger:2011ww, Maity:2019hre}). As in the present framework $S_2$ is unstable, after the breaking of its CE with $S_1$, $Y_2$ will continuously decrease as $x$ increases.}. As we focus on the coscattering, we do not include deviations induced by the contact terms of this Higgs portal scenario, therefore in the rest of the paper we assume they are sufficiently small to not deviate from the relic abundance calculation with \darkON.  

To end this section, we comment about the low mass regime $m_1 < m_h/2$, which turns out to be disfavored by LHC data. 
From the above discussion we found that coscattering requires nearly degenerate masses of the new scalars, i.e. $m_1 \lesssim m_2 < m_h/2$. On the other hand, we have checked numerically that in order to obtain the correct relic abundance $\lambda_{12}$ would be large enough to also recover CE within the dark sector, making coscattering ineffective unless $\lambda_{H2}\gtrsim 1$. However, searches of Higgs to invisible at the LHC have set limits on $\Gamma(h\rightarrow S_2S_2)$ \cite{ATLAS:2022yvh}, that translate into $\lambda_{H2} \lesssim 10^{-2}$. We have also checked that the inclusion of the contact terms does not change this result.

To summarize, we have presented the cBE for the system of $S_1$ and $S_2$, and we have solve them making use of the \micro \ code. The three regimes that we have distinguished, coscattering, mediator FO, and (co)annihilations, depend strongly on the parameters $\Delta m$ and $\lambda_{12}$, with coscattering favoring $\Delta m \ll (m_1,m_2)$ and $\lambda_{12} \ll 1$. Besides, on-shell (inverse) decay rates of $S_2$ are very efficient to maintain CE for much smaller values of $\lambda_{12}$ than in the case of off-shell decays. Contact terms are not essential to have coscattering, and we have seen that light DM is ruled-out by LHC bounds.

\section{Phenomenology}\label{sec3}
In this section we discuss direct detection and big-bang nucleosynthesis (BBN) constraints, and the prospects of having LLP in the coscattering scenario.

\subsection{Constraints}
\subsubsection{Direct detection}
The stable DM particle $S_1$ could be observed in direct detection experiments via the Higgs portal.
This implies a bound on the effective DM-Higgs coupling \cite{Cline:2013gha}
\begin{eqnarray}\label{eq:DD-constraint}
 \lambda_{H1} \lesssim \sqrt{\frac{4\pi m_h^4 m_1^2\sigma_{LZ}}{f_N^2 m_n^4 }},
\end{eqnarray}
where $\sigma_{LZ}$ denotes the upper bounds at 90\% C.L on the effective DM-nucleon scattering cross section obtained from the LZ experiment \cite{LZ:2022lsv}, $f_N\approx 0.3$ the effective nucleon-Higgs coupling, $m_n\approx 0.9$ GeV the nucleon mass, and $m_h = 125$ GeV the SM Higgs mass. 

As coscattering can be achieved for sufficiently small values of $\lambda_{H1}$, one could investigate the maximum values taken by this parameter without jeopardizing the relic abundance obtained by \darkON\, at the same time being in the ballpark of those values that yield a strong enough signal to be searched in future direct detection experiments. Certainly, $\lambda_{H1}$ can not take arbitrarily large values, otherwise CE is recovered by processes of the type $1 h \leftrightarrow 2 h$. To quantify the interplay among all these effects, in Fig.~\ref{plot4} we show the effect of $\lambda_{H1}$ on the relic abundance, for $m_1 = 90$ and $500$ GeV, $\Delta m = 1$ GeV and $\lambda_{H2} = 1$. Besides, the color indicates the value of $\Delta_{1s}^\Omega$ (see eq.~\ref{delta1s}). As expected, sizable values of $\lambda_{H1}$ decrease the relic abundance with respect to vanishing $\lambda_{H1}$, and for very large values of this parameter CE is recovered. However, LZ bounds (solid vertical lines) do not allow such sizable values of $\lambda_{H1}$, ruling out strong deviations from the co-scattering regime as given by \darkON\ . In particular, for $m_1 = 500$ GeV, it is possible to have sizable values of this parameter, i.e. $\lambda_{H1} \sim 10^{-2}$, in the ball park of LZ bounds (but still evading them), and without recovering CE. Actually, in that specific case, Darwin experiments \cite{DARWIN:2016hyl} will be sensitive to regions with even smaller values of $\lambda_{H1}$ (dashed vertical lines in Fig.~\ref{plot4}~(right)). 

 \begin{figure}[t!]
\centering
\includegraphics[width=0.55\textwidth]{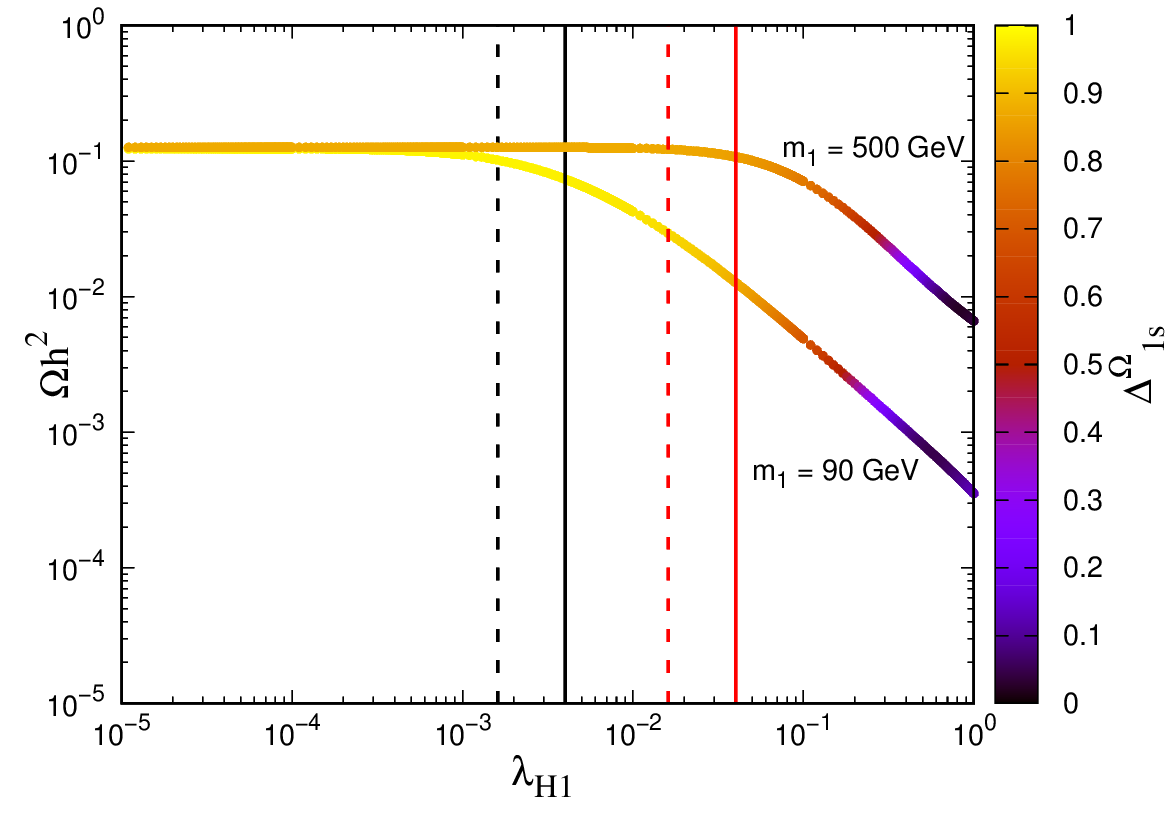}
\caption{Relic abundance as a function of $\lambda_{H1}$ considering $\Delta m = 1$ GeV, $\lambda_{H2} = 1$ and no contact terms. The vertical lines represent the bound from LZ \cite{LZ:2022lsv} and DARWIN projections \cite{DARWIN:2016hyl}, with the black solid (black dashed) and red solid (red dashed) lines representing the upper bounds for $m_1 = 90$ and 500 GeV, respectively. $\lambda_{12}$ has been chosen such that observed relic density is satisfied in the co-scattering regime, $\lambda_{12} = \{\ 3.86\times 10^{-4}, 2.3 \times 10^{-5}\}$ for $m_1 = \{90,500\}$ GeV respectively.}
\end{figure}

We point out that if loop corrections are considered, the DM-Higgs coupling
should be renormalized on-shell in order to retain agreement with eq.~\eqref{eq:DD-constraint}
at higher orders. We have outlined a suitable treatment of loop corrections in appendix \ref{sec:loop-corrections}
and found that in our model the loop effects to the observables of interest are negligible.

\subsubsection{BBN}
Additional stable or decaying particles present at temperatures $T\leq 10$ MeV may affect the measured primordial abundances of light elements. To our knowledge, constraints on lifetime of new  singlet scalars have been only considered for masses $\leq m_h/2$ \cite{Fradette:2017sdd}. We estimate the bounds coming from BBN using the results obtained in \cite{Jedamzik:2006xz}, considering the relic abundance of $S_2$ before its decay and the branching fraction of decays of $S_2$ into hadronic decays. In the present Higgs portal scenario, for $\Delta m \gtrsim 1$ GeV the model is practically safe of BBN constraints in the parameter space that we explore, since just after the decoupling of $S_2$ from the thermal plasma, $\Omega_{S_2} h^2$ is at least one order of magnitude below the measured DM relic abundance. The same conclusions were obtained in the leptophilic DM scenario in the coscattering mechanism \cite{Junius:2019dci}.

\subsection{Long-lived particles}
In the coscattering regime, the coupling $\lambda_{12}$ which determines the decay width of $S_2$ is very small, while simultaneously $\Delta m \ll (m_1, m_2)$. Therefore, the dark partner $S_2$ typically constitutes a long-lived particle (LLP) with a wide range of possible lifetimes in different regions of the parameter space. 
While single production of $S_2$ (like production of the DM relic $S_1$) at colliders is suppressed by $\lambda_{12}$, 
pair production of $S_2$ through an intermediate Higgs boson must be sizable, via the chain \cite{Craig:2014lda}
\begin{eqnarray}
    pp\rightarrow h^* + X \rightarrow S_2 + S_2 + X,
\end{eqnarray}
with $X$ being other states not relevant for the discussion. The goal of this section is to compare a few $S_2$ lifetime estimations predicted by the extended Higgs-singlet scenario that could be in the reach of present and future experiments, specially when the production mechanism is motivated by coscattering. In our knowledge, as LLP in Higgs portals have only been considered for mediator masses $\lesssim m_h/2$ \cite{Curtin:2018mvb, Alimena:2019zri}, the results presented here could motivate the search of heavier scalars through the Higgs portal. 

\begin{figure}[t!]
\centering
\includegraphics[width=1\textwidth]{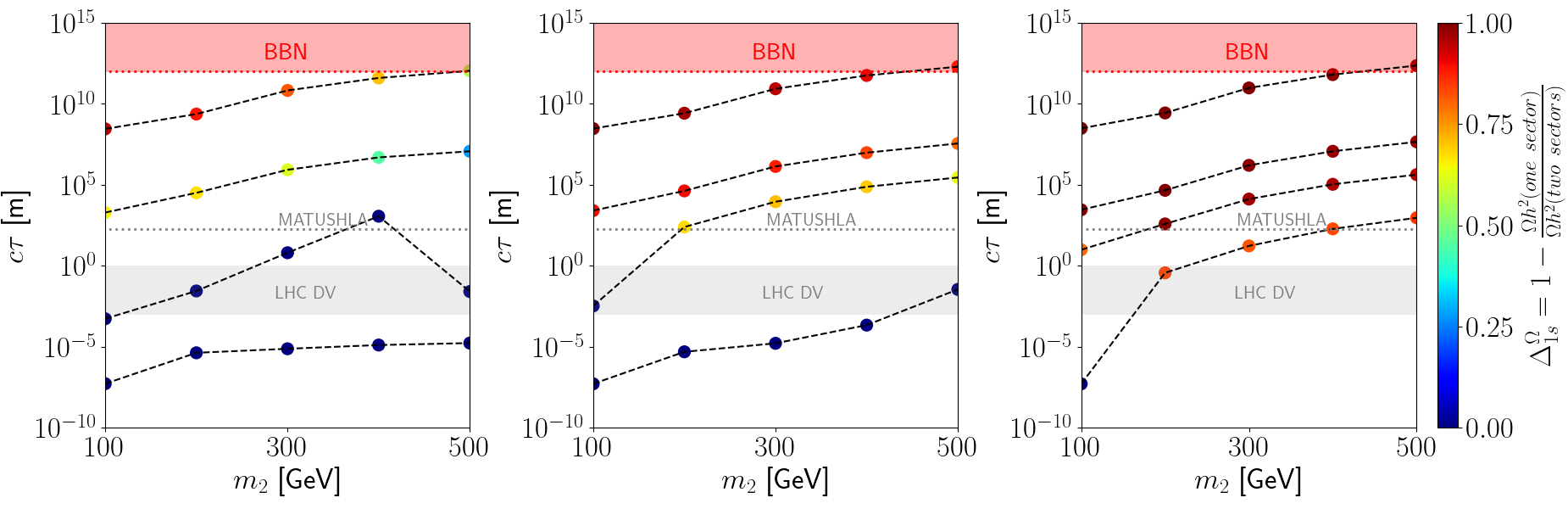}
\caption{Proper lifetime of the mediator as a function of its mass in the SBS, for $\lambda_{H2} = 0.5, 1$ and $\pi$, respectively, and $\lambda_{12}$ fixed to obtain the correct relic abundance. In each plot, from top to bottom $\Delta m = 1, 5, 10$ and 20 GeV, respectively. The color of each point represents the value of $\Delta_{1s}^\Omega$.}
\label{ctau}
\end{figure}

In Fig.~\ref{ctau}, we show the results for the lifetime of $S_2$ as a function of its mass for $\lambda_{H1}=0$ and 
fixed $\lambda_{H2} = 0.5$ (left), $1$ (middle) and $\pi$ (right). The row of points from top to bottom corresponds to 
$\Delta m =$ 1, 5, 10 and 20 GeV, while the color of each point indicates $\Delta_{1s}^\Omega$. 
The variation in $c\tau$ depends strongly on scalar mass difference, with small values of $\Delta  m$ favoring 
the coscattering regime ($\Delta_{1s}^\Omega \lesssim 1$), and in turn giving rise to enormous lifetimes of $S_2$, with some of the points well beyond earth size experiments, thereby confronting bounds coming from BBN. 
As $\Delta m$ increases, the values of $c\tau$ decrease to the point of reaching typical decay lengths for future
experiments such as MATHUSLA \cite{Curtin:2018mvb}. Notice that the reach of the latter may not only test particles that were produced in
the coscattering regime, but also probe the other two regimes that we studied in Sec~\ref{relicsub} (blue points). 
There are also model predictions in the reach of displaced vertex (DV) for ATLAS or CMS \cite{Lee:2018pag} (grey band in each plot). Finally, the blue
points in each plot are not unique for the corresponding chosen parameter space points shown in Fig.~\ref{ctau}, 
since as some of them belong to the mediator freeze-out regime (see region II in Fig.~\ref{plot2}), 
there is a range of $\lambda_{12}$ values fulfilling the measured relic abundance, then varying in orders of 
magnitude their corresponding $c\tau$ value.

\section{Conclusions}\label{sec4}
In this work, we have studied for the very first time the simplest Higgs portal scenario in the context of coscattering. This SM extensions considers two real scalars charged under a single $Z_2$ discrete symmetry, in which after EWSB, the lightest eigenstate is cosmologically stable, and the heavier one is unstable. We have explored in major detail the impact of each parameter in the thermal mechanism: coscattering, mediator freeze-out, and DM freeze-out. We put special attention to the first case, identifying parameter space for DM and mediator masses of hundreds of GeV giving the correct relic abundance. Radiative corrections do not generate significant deviations to the results that we obtain neither in direct detection nor in the calculation of the relic abundance. Besides, we have shown that the coscattering regime for the extended singlet-Higgs scenario gives rise to (very)long-lived mediators that could be in the reach of present and future experiments. Finally, effects of early kinetic decoupling \cite{Binder:2017rgn} on the relic calculation could modify at some extent the results presented in this work, but this analysis is beyond the scope of our work. 

\acknowledgments
We thank the creators and developers of \micro, and a special thank to Sasha Pukhov for his constant assistance on \micro. We thank Gael Alguero for providing important information relevant to our work. We also thank Giovanna Cottin for helpful advice on the long-lived particle subject. B.D.S has been founded by ANID (ex CONICYT) Grant No. 3220566. B.D.S. and J.L want to thank DESY and the Cluster of Excellence Quantum Universe, Hamburg, Germany, were this work was initiated.

\appendix
\section{Lagrangian original basis}\label{app1}
In this appendix we develop the details of the model in terms of the original field basis, which after some algebra becomes the simplified Lagrangian that we used in eq.~\ref{pot}.

Let us consider two real singlet scalars $\tilde{S}_1$ and $\tilde{S}_2$, charged under the same $Z_2$ symmetry such that $X \rightarrow X$ and $\tilde{S}_i \rightarrow -\tilde{S}_i$, with $i=1,2$ \cite{Casas:2017jjg, Ghorbani:2014gka}. The corresponding potential is given by
\begin{eqnarray}
 V(H, S_1, S_2) &\supset & \tilde{m}_1^2 \tilde{S}_1^2 + \tilde{m}_2^2 \tilde{S}_2^2 + \tilde{\lambda}_{H1}\tilde{S}_1^2 H^\dagger H + \tilde{\lambda}_{12}\tilde{S}_1\tilde{S}_2 H^\dagger H + \tilde{\lambda}_{H2}\tilde{S}_2^2 H^\dagger H \nonumber\\   &+& \tilde{\lambda}_{22}\tilde{S}_1^2 \tilde{S}_2^2 + \tilde{\lambda}_{13}\tilde{S}_1 \tilde{S}_2^3 + \tilde{\lambda}_{31}\tilde{S}_1^3 \tilde{S}_2.
\label{pot1}
\end{eqnarray}
After EWSB, with $\ev{H} = (0 , v_h)^T/\sqrt{2}$, the scalars $\tilde{S}_1$ and $\tilde{S}_2$ mix, but after rotation the potential can be written identically as in \ref{pot1}. We diagonalize using 
\begin{eqnarray}\label{diago}
  O^T\mathcal{M}^2O = \text{diag}(m_1^2,m_2^2)
\end{eqnarray}
The mass matrix is
\begin{eqnarray}
 \mathcal{M}^2 = \begin{pmatrix}
\tilde{m}_1^2 + \tilde{\lambda}_{H1}v_h^2 & \tilde{\lambda}_{12} v_h^2/2\\
\tilde{\lambda}_{12} v_h^2/2 & \tilde{m}_2^2 + \tilde{\lambda}_{H2}v_h^2
\end{pmatrix}
\end{eqnarray}
The Eigenmass are given by
\begin{eqnarray}
 m_1^2 &=& (\tilde{m}_1^2 + \tilde{\lambda}_{H1}v_h^2)\cos^2\theta + (\tilde{m}_2^2 + \tilde{\lambda}_{H2}v_h^2)\sin^2\theta - \sin\theta\cos\theta \tilde{\lambda}_{12}v_h^2  \\ 
 m_2^2 &=& (\tilde{m}_2^2 + \tilde{\lambda}_{H2}v_h^2)\cos^2\theta + (\tilde{m}_1^2 + \tilde{\lambda}_{H1}v_h^2)\sin^2\theta + \sin\theta\cos\theta \tilde{\lambda}_{12}v_h^2
\end{eqnarray}
Additionally, from the non-diagonal relationship of eq.~\ref{diago} we obtain that 
\begin{eqnarray}\label{relation}
 \tan(2\theta) = -\frac{\tilde{\lambda}_{12}v_h^2}{2(\tilde{m}_1^2 - \tilde{m}_2^2 + v_h^2(\tilde{\lambda}_{H1} - \tilde{\lambda}_{H2}))}
\end{eqnarray}
Replacing eq.~\ref{relation} into the original Lagrangian, and writing down eq.~\ref{pot1} in terms of the physical states, i.e. $\tilde{S}_1 = \cos\theta S_1+ \sin\theta S_2$ and $\tilde{S}_2 = -\sin\theta S_1+ \cos\theta S_2$, we obtain the potential presented in eq.~\ref{pot}.

\section{Dark Matter Conversion Rate}\label{App:conversion-rate}
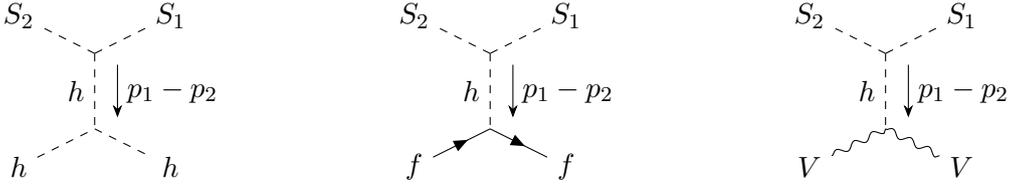
\begin{figure}
    \centering
    \begin{tikzpicture}[baseline=-.1em, scale=.5]
        \begin{feynman}[small]
            \vertex (S2) at (-2,2) {$S_2$};
            \vertex (S1) at (2, 2) {$S_1$};
            \vertex (f1) at (-2,-2) {$h$};
            \vertex (f2) at (2,-2) {$h$};
            \vertex (a) at (0, 1);
            \vertex (b) at (0,-1);
            \diagram* {
                (S1) -- [scalar] (a) -- [scalar] (S2),
                (a) -- [scalar, edge label'=$h$, momentum=$p_1-p_2$] (b),
                (f1) -- [scalar] (b) -- [scalar] (f2)
            };
        \end{feynman}
    \end{tikzpicture}\hspace{2cm}
    \begin{tikzpicture}[baseline=-.1em, scale=.5]
        \begin{feynman}[small]
            \vertex (S2) at (-2,2) {$S_2$};
            \vertex (S1) at (2, 2) {$S_1$};
            \vertex (f1) at (-2,-2) {$f$};
            \vertex (f2) at (2,-2) {$f$};
            \vertex (a) at (0, 1);
            \vertex (b) at (0,-1);
            \diagram* {
                (S1) -- [scalar] (a) -- [scalar] (S2),
                (a) -- [scalar, edge label'=$h$, momentum=$p_1-p_2$] (b),
                (f1) -- [fermion] (b) -- [fermion] (f2)
            };
        \end{feynman}
    \end{tikzpicture}\hspace{2cm}
    \begin{tikzpicture}[baseline=-.1em, scale=.5]
        \begin{feynman}[small]
            \vertex (S2) at (-2,2) {$S_2$};
            \vertex (S1) at (2, 2) {$S_1$};
            \vertex (f1) at (-2,-2) {$V$};
            \vertex (f2) at (2,-2) {$V$};
            \vertex (a) at (0, 1);
            \vertex (b) at (0,-1);
            \diagram* {
                (S1) -- [scalar] (a) -- [scalar] (S2),
                (a) -- [scalar, edge label'=$h$, momentum=$p_1-p_2$] (b),
                (f1) -- [boson] (b) -- [boson] (f2)
            };
        \end{feynman}
    \end{tikzpicture}
    \caption{Tree-level contributions to $S_2\to S_1$ conversions in the thermal bath.}
    \label{fig:conversion-diagrams}
\end{figure}
In this section we present a calculation of the thermally averaged cross section for DM conversion $\ev{\sigma_{2X\to 1X} v}$ where $X$ can be any SM particle. 
The differential cross section of the conversion process in the centre of mass frame is given by
\begin{align}
    \left(\frac{d\sigma_{2X\to 1X} v}{d\Omega}\right)_{c.o.m.} = \frac{|\mathbf{p_f}|}{64\pi^2 E_2 E_X \sqrt{s}} \overline{|\mathcal{M}|}_{2X\to 1X}^2,
\end{align}
where $\sqrt{s}$ denotes the total c.o.m. energy, $|\mathbf{p_f}|=\lambda(s,m_1^2,m_X^2)^{1/2}/(2\sqrt{s})$ denotes the final state momentum\footnote{$\lambda(x,y,z)=(x-y-z)^2-4yz$ is the Källén function.} and
$E_2=(|\mathbf{p_i}|^2+m_2^2)^{1/2}$ and $E_X=(|\mathbf{p_i}|^2+m_X^2)^{1/2}$ denote the energies of the initial state DM and SM particle 
with momentum $|\mathbf{p_i}|= \lambda(s,m_2^2,m_X^2)^{1/2}/(2\sqrt{s})$. At tree-level the conversion processes are possible for $X=h,f,W,Z$ through the $t$-channel 
diagrams shown in Fig. \ref{fig:conversion-diagrams}. The resulting squared matrix elements are given by
\begin{subequations}
    \begin{align}
        \overline{|\mathcal{M}|}_{2h\to 1h}^2 &= \frac{9\lambda^2_{12} m_h^4}{(t-m_h^2)^2},\\
        \overline{|\mathcal{M}|}_{2f\to 1f}^2 &= \frac{2\lambda^2_{12} m^2_f(4m_f^2 - t)}{(t-m_h^2)^2},\\
        \overline{|\mathcal{M}|}_{2V\to 1V}^2 &= \frac{4\lambda^2_{12} m_V^2 (3 m_V^2 + t)}{(t-m_h^2)^2},
    \end{align}
\end{subequations}
where 
\begin{align}
    t = (p_1-p_2)^2 = 2 m_X^2 - 2 (|\mathbf{p_i}|^2+m_X^2)^{1/2} (|\mathbf{p_f}|^2+m_X^2)^{1/2} + 2 |\mathbf{p_f}||\mathbf{p_i}| \cos\theta.
\end{align}
After substitution, the solid angle differential becomes $d\Omega = d\varphi dt / (2 |\mathbf{p_f}||\mathbf{p_i}|)$ and the integrals can be solved
to obtain the total cross section
\begin{subequations}
    \begin{align}
        \sigma_{2h\to 1h} v &= \frac{9 \lambda_{12}^2 m_h^4 |\mathbf{p_f}|}{16\pi E_2 E_X\sqrt{s}} \frac{1}{(m_h^2-t^-)(m_h^2-t^+)},\\
        \sigma_{2f\to 1f} v &= \frac{\lambda_{12}^2 m_f^2}{32\pi E_2 E_X |\mathbf{p_i}|\sqrt{s}} \bigg[ \ln(\frac{m_h^2-t^-}{m_h^2-t^+}) 
        - \frac{4|\mathbf{p_i}||\mathbf{p_f}|(m_h^2-4m_f^2)}{(m_h^2-t^-)(m_h^2-t^+)}  \bigg], \\
        \sigma_{2V\to 1V} v &= \frac{\lambda_{12}^2 m_V^2}{16\pi E_2 E_X |\mathbf{p_i}|\sqrt{s}} \bigg[ 
        \frac{4|\mathbf{p_i}||\mathbf{p_f}|(m_h^2+3m_V^2)}{(m_h^2-t^-)(m_h^2-t^+)} - \ln(\frac{m_h^2-t^-}{m_h^2-t^+}) \bigg],
    \end{align}
\end{subequations}
where $t^\pm\equiv t(\cos\theta=\pm 1)$. Next, the thermal average has to be calculated from
\begin{align}
    \ev{\sigma v} = \int_{(m_2+m_X)^2}^\infty \frac{E_2 E_X \sigma v}{4 m_2^2 m_X^2 T} 
    \frac{K_1({\textstyle\frac{\sqrt{s}}{T}})}{K_2({\textstyle\frac{m_2}{T}})K_2({\textstyle\frac{m_X}{T}})} 
    \sqrt{s - 2(m_2^2 + m_X^2) + \frac{(m_2^2-m_X^2)^2}{s}}.
\end{align}
To good approximation, this integral is given by $\ev{\sigma v} \approx \sigma v(s=\ev{s})$ where 
\begin{align}
    \ev{s} \approx (m_2+m_X)^2 + 6 (m_2+m_X) T  + \mathcal{O}(T^2/m_2^2),
\end{align}
such that the thermally averaged cross sections, in the limit
$T,\Delta m \ll m_{1,2}$, are given by
\begin{subequations}
    \begin{align}
    \ev{\sigma_{2h\to 1h} v} &\approx \frac{9\lambda_{12}^2}{8\pi}
    \sqrt{\frac{\Delta m + 3T}{2 m_2 m_h(m_2+m_h)^3}} \\
    \ev{\sigma_{2f\to 1f} v} &\approx \frac{\lambda_{12}^2 m_f^4}{\pi m_h^4} \sqrt{\frac{\Delta m + 3T}{2 m_2 m_f (m_2+m_f)^3}} \\
    \ev{\sigma_{2V\to 1V} v} &\approx \frac{3\lambda_{12}^2 m_V^4}{2\pi m_h^4} \sqrt{\frac{\Delta m + 3T}{2 m_2 m_V (m_2+m_V)^3}}
\end{align}
\end{subequations}
Finally, the DM conversion rate is $\Gamma_{1\to 2} = (n_2^e/n_1^e)\sum_{X} \ev{\sigma_{2X\to 1X}v} n_X^e$.

\section{Treatment of Radiative Corrections}\label{sec:loop-corrections}

\begin{figure}[h]
	\centering
	\begin{tikzpicture}[baseline=-.5ex, scale=.8]
		\begin{feynman}[small]
			\vertex (s1) at (-1.5,1) {$DM$};
			\vertex (s2) at (1.5, 1) {$DM$};
			\node[blob] (a) at (0, 0.5) {$\lambda$}; 
			\node[dot] (b) at (0, -0.5);
			\vertex (x1) at (-1.5,-1) {$SM$};
			\vertex (x2) at (1.5,-1) {$SM$};
			\diagram*{
				(s1) -- [scalar] (a) -- [scalar] (s2),
				(a) --[scalar, edge label' = $h$,momentum={$q^2=0$}] (b),
				(x1) -- [double] (b) -- [double] (x2)
			};
		\end{feynman}
	\end{tikzpicture}\hspace{2cm}
	\begin{tikzpicture}[baseline=-.5ex, scale=.8]
		\begin{feynman}[small]
			\vertex (s1) at (-1.5,1) {$DM$};
			\vertex (s2) at (-1.5, -1) {$DM$};
			\node[blob] (a) at (-0.5, 0) {$\lambda$}; 
			\node[dot] (b) at (0.5, 0);
			\vertex (x1) at (1.5, 1) {$SM$};
			\vertex (x2) at (1.5,-1) {$SM$};
			\diagram*{
				(s1) -- [scalar] (a) -- [scalar] (s2),
				(a) --[scalar, edge label' = $h$,momentum={$q^2=\ev{s}$}] (b),
				(x1) -- [double] (b) -- [double] (x2)
			};
		\end{feynman}
	\end{tikzpicture}
	\caption{Schematic diagrams showing the contribution of the effective DM-Higgs vertex $\lambda_{DM}(q^2)$ to direct detection (left) and coannihilation processes during freeze-out (right). }
    \label{fig:effective-coupling}
\end{figure}
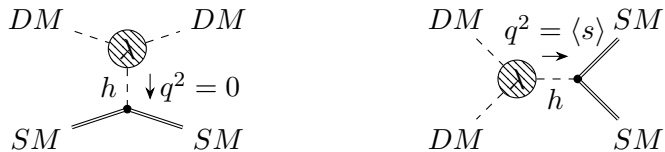

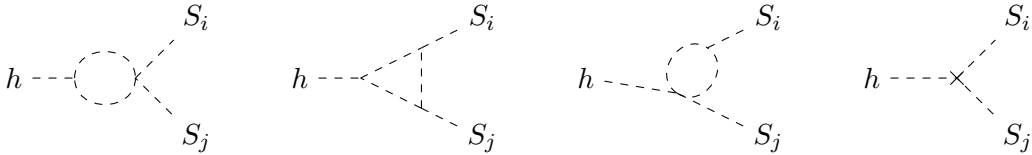
\begin{figure}
	\centering
	\begin{tikzpicture}[baseline=-.5ex, scale=.8]
		\begin{feynman}[small]
			\vertex (h) at (-2,0) {$h$};
			\vertex (s1) at (1, 1) {$S_i$};
			\vertex (s2) at (1, -1) {$S_j$};
			\vertex (a) at (0, 0); 
			\vertex (b) at (-1,0);
			\diagram*{
				(h) -- [scalar] (b) -- [scalar, half left] (a) -- [scalar] (s1),
				(s2) -- [scalar] (a) -- [scalar, half left] (b)
			};
		\end{feynman}
	\end{tikzpicture} \hfil
        \begin{tikzpicture}[baseline=-.5ex, scale=.8]
		\begin{feynman}[small]
			\vertex (h) at (-2,0) {$h$};
			\vertex (s1) at (1, 1) {$S_i$};
			\vertex (s2) at (1, -1) {$S_j$};
			\vertex (a) at (0, 0.5); 
			\vertex (b) at (-1,0);
                \vertex (c) at (0, -0.5);
			\diagram*{
				(h) -- [scalar] (b) -- [scalar] (a) -- [scalar] (s1),
				(s2) -- [scalar] (c) -- [scalar] (b),
                    (c) -- [scalar] (a)
			};
		\end{feynman}
	\end{tikzpicture} \hfil
        \begin{tikzpicture}[baseline=-.5ex, scale=.8]
		\begin{feynman}[small]
			\vertex (h) at (-2,0) {$h$};
			\vertex (s1) at (1, 1) {$S_i$};
			\vertex (s2) at (1, -1) {$S_j$};
			\vertex (a) at (0, 0.5); 
			\vertex (b) at (-0.5,-0.25);
			\diagram*{
				(h) -- [scalar] (b) -- [scalar, half left] (a) -- [scalar] (s1),
				(s2) -- [scalar] (b) -- [scalar, half right] (a)
			};
		\end{feynman}
	\end{tikzpicture} \hfil
	\begin{tikzpicture}[baseline=-.5ex, scale=.8]
		\begin{feynman}[small]
			\vertex (h) at (-1.4,0) {$h$};
			\vertex (s1) at (1, 1) {$S_i$};
			\vertex (s2) at (1, -1) {$S_j$};
			\vertex (a) at (0, 0); 
			\diagram*{
				(h) -- [scalar, insertion=1] (a) -- [scalar] (s1),
				(s2) -- [scalar] (a)
			};
		\end{feynman}
	\end{tikzpicture}
	\caption{Diagrams contributing to $h\to S_i S_j$ at one-loop order. }
	\label{fig:one-loop-diagrams}
\end{figure}

In this appendix we outline the on-shell renormalization of the model and estimate the impact of one-loop corrections 
on the results obtained in this paper. We note that a proper definition of the renormalization conditions is crucial
in order to obtain meaningful results at NLO. In particular, the physical interpretation
of the parameters at tree-level is only retained if they fulfill corresponding on-shell renormalization conditions at 
one-loop order. In other schemes like $\overline{MS}$ or for ad-hock subtractions the model parameters no longer correspond 
directly to the observables of interest. The parameters relevant for a renormalization of
the scalar sector are the scalar masses $m_i^2$, quartic couplings $\lambda_{ij}$ and Higgs vev $v_h$. The renormalized 
Lagrangian is obtained from the following renormalization transformation of the bare parameters
\begin{align}
    m_{i,0}^2 \to m^2_{i} + \delta m^2_i, \qquad \lambda^0_{ij} \to \lambda^R_{ij} + \delta \lambda_{ij}, \qquad
    v_h^0 \to v_h + \delta v_h,
\end{align}
and renormalization of the bare fields
\begin{align}
    S_i^0 \to \sqrt{Z_i} S_i, \qquad H_0 \to \sqrt{Z_H} H, \qquad \text{where} \quad 
    Z_i = 1 + \delta Z_i.
\end{align}
After on-shell renormalization, the scalar masses $m_i$ 
correspond to the physical pole masses of 
the DM particles and the scalar couplings $\lambda^R_{ij}$ correspond to physical effective
coupling strengths measured e.g. in direct detection or collision experiments. This implies
a set of conditions on the corresponding amplitudes from which the renormalization constants can be 
determined. Here, we demonstrate this specifically for $\lambda_{H1}$ and 
$\lambda_{12}$, which where chosen to be very small in the above analysis. 
We define $\lambda_{H1}$ to be the effective $S_1$-Higgs coupling measured in the direct detection experiments sketched in Fig. \ref{fig:effective-coupling} (left), while $\lambda_{12}$
is defined through DM production and annihilation events at $\ev{s}=(m_1+m_2)^2$ such as in 
Fig. \ref{fig:effective-coupling} (right). Note that, in general, the effective (quantum corrected) DM-Higgs couplings denoted by $\lambda_{ij}(q^2)$ will be dependent on the (off-shell) Higgs momentum $q$. At one-loop these effective couplings are given by
\begin{subequations}
    \begin{align}
        \lambda_{H1}(q^2) &= \lambda^R_{H1} + \Gamma_{H1}(q^2) + \delta\lambda_{H1} 
        + \lambda^R_{H1} \Big(\delta Z_1 + \tfrac{1}{2}\delta Z_H + \frac{\delta v_h}{v_h} \Big) \\
        \lambda_{12}(q^2) &= \lambda^R_{12} + \Gamma_{12}(q^2) + \delta\lambda_{12}
        + \lambda_{12}^R\Big(\frac{1}{2}\delta Z_1 + \frac{1}{2}\delta Z_2 + 
        \frac{1}{2}\delta Z_H + \frac{\delta v_h}{v_h}\Big)
    \end{align}
\end{subequations}
where $\Gamma_{ij}(q^2)$ denotes the contributions of the one-loop diagrams from Fig. 
\ref{fig:one-loop-diagrams}. The definitions of the coupling strenghts translate into the following renormalization conditions
\begin{align}
    \lambda_{H1}^R \equiv \lambda_{H1}(0), \qquad \lambda_{12}^R
    \equiv \lambda_{12}(\ev{s})
\end{align}
And can easily be fulfilled by choosing the renormalization constants
$\delta \lambda_{H1}$ and $\delta \lambda_{12}$ appropriately. In the limit $\lambda^R_{H1},\lambda^R_{12}\approx0$ the only contributing one-loop diagrams are of the type Fig. \ref{fig:one-loop-diagrams} (left) and
result in the following expressions for the renormalized vertex functions 
\begin{align}
    \lambda_{H1}(q^2) &= \lambda^R_{H1} -\frac{\lambda^R_{22}\lambda^R_{H2}}{8\pi^2} \Big(B_0(q^2,m_2^2,m_2^2) - B_0(0,m_2^2,m_2^2)\Big)\\
    \lambda_{12}(q^2) &= \lambda^R_{12} -\frac{3\lambda^R_{13}\lambda^R_{H2}}{16\pi^2} \Big(B_0(q^2,m_2^2,m_2^2) - B_0(\ev{s},m_2^2,m_2^2)\Big)
\end{align}
By definition of the on-shell scheme, quantum corrections to direct detection of $S_1$ and to annihilation and production processes of $S_1+S_2$ during freeze-out are 0. Corrections only appear for annihilation and production of $S_1$, where the relevant momentum transfer is $q^2=4 m_1^2$. The resulting effective
coupling that should be used when calculating the relic abundance is (for $m_1\simeq m_2$)
\begin{align}
	\lambda_{H1}(4m_1^2) \approx \lambda^R_{H1} -\frac{\lambda^R_{22}\lambda^R_{H2}}{4\pi^2}
\end{align}
The loop corrections, as expected, result in very small $\mathcal{O}(1\%)$ effects and do not have any
important impact on the above analysis.

\bibliographystyle{JHEP}
\bibliography{bibliography}

\end{document}